\documentclass[10pt,journal]{IEEEtran}
\usepackage{ifpdf}

\ifpdf
 \usepackage{graphicx}

 \DeclareGraphicsExtensions{.pdf,.jpeg,.png}
\else
 \usepackage{graphicx}

\DeclareGraphicsExtensions{.eps}
\fi

\usepackage{epstopdf}
\usepackage{epsfig}

\usepackage{algorithmic}
\usepackage{algorithm}
\usepackage[font=normalsize]{caption}
\usepackage[tight,normalsize]{subfigure}
\begin{document}
%
\title{Adaptive Streaming in P2P Live Video Systems: \\
A Distributed Rate Control Approach}

\author{\IEEEauthorblockN{Laura Natali and Maria Luisa Merani\\}
\IEEEauthorblockA {Dipartimento di Ingegneria ``Enzo Ferrari''\\
University of Modena and Reggio Emilia, Modena, Italy\\
e-mail: laura.natali@unimore.it, marialuisa.merani@unimore.it}}


%


\maketitle

\begin{abstract}

Dynamic Adaptive Streaming over HTTP (DASH) is a recently proposed standard that offers different versions of the same media content to adapt the  delivery process over the Internet to dynamic  bandwidth fluctuations and different user device capabilities. 
The peer-to-peer (P2P) paradigm for video streaming allows to leverage  the cooperation among peers, guaranteeing  to serve every video request with increased scalability and reduced cost. 
We propose to combine these two approaches in a P2P-DASH architecture, exploiting the potentiality of both. 
The new platform is made of several swarms, and a different DASH representation is streamed within each of them; unlike client-server DASH architectures, where each client autonomously selects which version to download according to current network conditions and to its device resources, we put forth a new rate control strategy implemented at peer site to maintain a good 
viewing quality to the local user 
and to simultaneously guarantee the successful operation of the P2P swarms.
The effectiveness of the solution is demonstrated through simulation and it indicates 
that the P2P-DASH platform is able to warrant its users a very good performance, 
much more satisfying than in a conventional P2P environment where DASH is not employed. 
Through a comparison with a reference DASH system modeled via the Integer Linear Programming (ILP) approach, 
the new system is shown to outperform such reference architecture.  
To further validate the proposal, both in terms of robustness and scalability, system behavior is  investigated in the critical condition of a flash crowd, showing that the strong upsurge of new users can be successfully revealed and gradually accommodated.
\end{abstract}


{\emph{keywords:} Adaptive streaming, DASH, peer-to-peer, flash-crowd, integer linear programming}

%
\IEEEpeerreviewmaketitle

\section{Introduction}
\label{sec:intro}

As several studies related to the use
of the Internet indicate \cite{CISCO} \cite{EUAGENDA}, we are witnessing
an explosive increase of video content; IP video flows
currently represent more than $50\%$ of all Internet traffic and are expected
to grow further at a very swift pace. End-users consuming video
typically rely upon different devices, ranging from smartphones
to tablets and PCs, and also undergo
different network connectivity conditions on
the basis of their location: at home, in office, on the road.
To  counteract fluctuating network conditions and also to cope
with heterogeneous user requests,
a new delivery
streaming framework termed dynamic adaptive streaming has been recently introduced.
Several proprietary solutions have first flourished, ranging from Smooth
Streaming by Microsoft \cite{MSSTREAMING} \cite{ZAMBELLI} to HTTP Live Streaming by Apple \cite{APPLESTREAMING} and
Dynamic HTTP Streaming by Adobe \cite{ADOBESTREAMING}.
They have in turn triggered the release of the 
Dynamic Adaptive Streaming over HTTP
(DASH) standard \cite{IFDASH}\cite{IEEEDASH},
which is currently the most widespread, internationally recognized approach: in DASH, the ``adaptive'' term refers to the capability of the technique to
modify the features of the transmitted video, in order to adapt it to varying network conditions and different user requests. 

This work puts forth a P2P system that exploits DASH technology for video streaming: in doing so, it embraces the future vision of the Internet provided in \cite{BABA}, where network users are seen as collectively forming the ``human grid'', whose distributed resources are at the basis of any service provisioning.
A multi-overlay architecture is therefore proposed where the  role of the servers in the distribution process is minimal; moreover, there are as many overlays as the number of available DASH representations and
an entirely distributed control strategy rules the transition of the peers  from one overlay to another.
Such control logic is locally implemented at the peer's site and it strives to preserve both the quality that the
single peer experiences and the good functioning of all overlays.
In order to jointly achieve these goals, it relies upon local and global parameters: the former reflect the peer's status; the latter indicate the status of the different overlays.
System behavior is explored in some relevant user scenarios 
and the findings are summarized as follows:
\begin{itemize}
\item  the new distributed control allows the pure P2P platform to
successfully reach steady-state, providing the majority of its users with the
desired video representation; 
\item in the most unfriendly scenario that has been investigated, the performance that the peers experience is significantly better than the one they would undergo in a DASH-unaware P2P system; 
\item when compared to the  performance that a reference, idealized system would achieve, in the most favorable setting the current proposal attains a very close number 
of satisfied users, i.e., users that stream the video at the desired bit rate,  
whereas  in the worst case,  the number of satisfied users is slightly suboptimal, but to the advantage of  
a much better streaming quality;
\item the critical upsurge of a flash crowd is promptly revealed by the control algorithm and is satisfyingly handled;
\item combining the DASH feature with P2P offers a significant advantage in terms of switching delay that the peers experience when requesting different video alternatives.
\end{itemize}

The rest of the paper is organized as detailed below: 
Section II critically provides the state of the art and covers the related work; Section III illustrates the proposed architecture and the distributed rate control algorithm, while
Section IV describes the integer linear programming approach that contributes to validate the proposal.
In Section V a numerical picture of the performance attained
by the new platform is offered and Section VI draws the conclusions.

\section{State of the Art}
The majority of the works on DASH, the new adaptive streaming standard, 
focuses on client-server architectures and in this setting, different algorithms to  
dynamically choose the most appropriate DASH representation
have been proposed. 
Among some recently published contributions,
the study in \cite{DASH-CONTROLTHEORY} investigates  
the adoption of DASH 
over multiple content distribution servers. In the examined architecture, the rate adaptation logic 
is applied to blocks of video fragments,   
downloaded in parallel from different servers;
a  novel  proportional-derivative  controller is further introduced to  adapt
the video bit rate, and system performance
is studied  through
a theoretical approach
and some  laboratory experiments.
In a similar manner, in a P2P architecture each user in parallel requests missing video content to several neighbors,  that is, each of them acts as a concurrent server;  in this context however, the pool of  peers  that provide useful content  varies over time, often unpredictably.

Various bandwidth-based solutions have also  been introduced in order to perform rate control in DASH, as 
in \cite{DASH-RATECONTROL}, that suggests
an adaptation algorithm relying on the smoothed Http throughput  
to determine if the current media bit rate 
matches the end-to-end network bandwidth,
or
as in \cite{DASH_bandwidth_based}, whose  strategy
relies upon the TCP download throughput of the client to determine 
the available bandwidth in the presence of congestion, 
whereas it constantly probes the network
and adapts to its new conditions otherwise.
The status of the playback buffer is considered in \cite{QDASH}, that  strives to
preserve the minimum buffer length that avoids interruptions and minimizes 
video quality variations during the playback and so does 
\cite{DASH-RATECONTROL2}, whose authors investigate a
more conservative, QoE-driven scheme.
More generally, current network conditions are monitored to efficiently -- and hopefully fairly --  leverage network resources, whereas the status of the client's buffer is examined to guarantee a smooth playout and avoid video stalls, the focal point being  the client's streaming quality.
%
Yet, the algorithms that govern the video streaming requests are steered only by the user's perspective:
that is, the client's  objective is to optimize its viewing experience, regardless of  the impact that its decisions have on the video quality perceived by other users.
To the authors' knowledge, in the client-server strategies
proposed in literature 
the user is not directly interested to the status of the whole system and at the same time it does not possess the concepts of cooperation or partnership with other users.
%
%

The present proposal employs an opposite approach: in its decision taking process, the peer takes into account both its local point of view and an overall perspective;
as a matter of fact, in a P2P architecture, where video streaming is achieved owing to the collaboration of all nodes, selfish behaviors do not produce advantage either to the single node or to the whole system.
Hence, a generic peer cannot neglect the status of the nodes within its swarm to attain a satisfying video quality. 
%
%
Differently from the previous works, we therefore propose a new rate control algorithm that strives to guarantee a good performance to the single user without loosing the overall system sight: 
hence, a node selects the most proper DASH representation taking into account some local, indirect video quality measures and also some parameters that reflect  the health status of the P2P system.
%
%
%
%

The contributions in \cite{DASH-P2PASSISTED} and \cite{ROSS1} are worth noting, as they propose the joint adoption of P2P and DASH. The former work proposes a standard compliant solution to integrate peer-assisted streaming in conventional DASH client-server systems, and
represents
a suitable approach in a
CDN (Content Delivery Network) type of environment; the latter
applies DASH technique to a P2P architecture  and 
exploits game theory 
to rule the switching process among different representations.
Differently from \cite{DASH-P2PASSISTED} and \cite{ROSS1}, we do not examine a Video on Demand
(VoD) system:  both works correctly investigate
platforms whose population size is very limited, as it is 
the case for VoD,
and interpret
P2P as a secondary, although beneficial, feature of the video 
distribution architecture; on the contrary, our study delves the 
feasibility and the achievable performance of
a system that delivers a \emph{live streaming
channel} to a \emph{large floor of users}, with  \emph{very little contribution from the
server}.

\section{P2P-DASH}
\label{sec:P2P-DASH}

\subsection{DASH Essentials}
Dynamic Adaptive Streaming over Http is a  
standard that serves the purpose of delivering multimedia content through the network via conventional web servers. 
In DASH, the different constituent components of the multimedia content, that is, 
video, audio and text, are made
available to the final consumer in different versions, displaying
different characteristics: as an example, these might be alternative encoded versions at different bit rates of the audio and video part, or different language versions of the audio content  in the case of pre-recorded material. 
The multimedia content is interpreted by DASH as a sequence of consecutive 
temporal periods: 
a manifest file, 
the Multimedia Presentation Description (MPD),
details the alternatives -- also termed \emph{representations} -- that are available for each component in every period; as the content of the periods is accessed as a collection of 
media segments, the MPD also provides the url where these segments can be accessed at.
The MPD is initially passed to the DASH client, 
which next proceeds to issue Http  get requests to retrieve the segments of the desired representations; during the streaming session, the client can  
dynamically adjust the client-sever communication, moving from one representation
to another.
The DASH standard does not indicate which adaptation algorithm to employ, nor
it details the fetching mechanism that the client adopts, that are left entirely open.
  
\subsection{P2P System Architecture}
In our proposed architecture P2P is the significant feature that rules the distribution of multimedia content: users are therefore organized in a virtual network, termed overlay, which is built at application layer, and are actively involved in the video diffusion process. They relay portions of the content they already own in their local buffer to other users - their peers - through the virtual links of the overlay. We consider the case where a single video channel has to be streamed to a 
population of users whose size is $N$  and assume that $K$ DASH representations of the video are instantaneously 
available: unlike
VoD client-server architectures, where $K$ takes on relatively high values, in
the examined scenario $K$ is intentionally confined to lower values, not to excessively complicate  
the design of a live system which has to respect tighter latency constraints.

\begin{figure}[h]
\centering
\includegraphics[trim = 0mm 0mm 0mm 0mm, clip,width= \columnwidth]{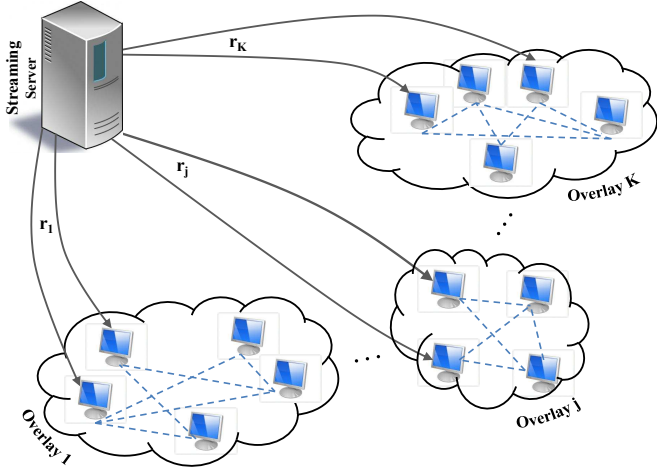}
\caption{System architecture} 
\label{fig:architecture}
\end{figure}

We assume that the bit rate is what distinguishes the different available DASH representations, each being distributed within a separate P2P overlay: 
the $j$-th representation exhibits a streaming rate $r_j$ and is delivered 
within the $j$-th swarm, $1 \le j \le K$; without loss of generality, we set 
$r_{j}<r_{j+1}$, $\forall j$, $j=1, 2, \ldots, K-1$.
As Fig.\ref{fig:architecture} indicates, the server is responsible 
for initially igniting the diffusion process: it 
delivers each 
representation to the corresponding overlay as a sequence of  
video chunks, relatively small fragments of the encoded video.
Every peer belongs to one overlay at a time and contributes to sustain the process with its own upload bandwidth, further spreading the video chunks that it receives to other peers belonging to the same overlay.
Peers have heterogeneous upload and download capacities and therefore supply content
to other mates in a diversified manner. 

The examined P2P overlays are mesh-based, and within each of them
every peer exchanges video chunks with $M$ randomly chosen neighbors, implementing a pull-based  protocol.
The peer periodically 
informs its neighbors about the
video chunks that it holds in its streaming buffer by forwarding  its buffer maps.
Moreover, the peer periodically  asks its neighbors for the chunks that are missing within its current
request window $W$. Such window identifies the range where
chunks that can be requested by the node fall: its right edge coincides
with the highest chunk sequence number that the node is aware of from the
inspection of the buffer maps it received, its left edge is $X$ s behind. The
window slides forward whenever the peer learns through the new buffer 
maps that it receives from its neighbors that a chunk with a higher sequence number is available; when a chunk
that is yet to be received falls outside the current request window, it will not
be claimed any longer, as outdated.

The DASH representation that a peer desires is the representation featuring the streaming rate 
the peer aims at viewing the 
video content at: we observe that it depends on the combined limits set by
the user's access network and by the display characteristics of 
its terminal;
the user's preferences might also contribute to determine it, as it is the
case when different charging rates
are applied for distinct video qualities. 

\begin{figure}[h]
\centering
\includegraphics[trim = 10mm 50mm 15mm 50mm, clip,width= 0.8\columnwidth]{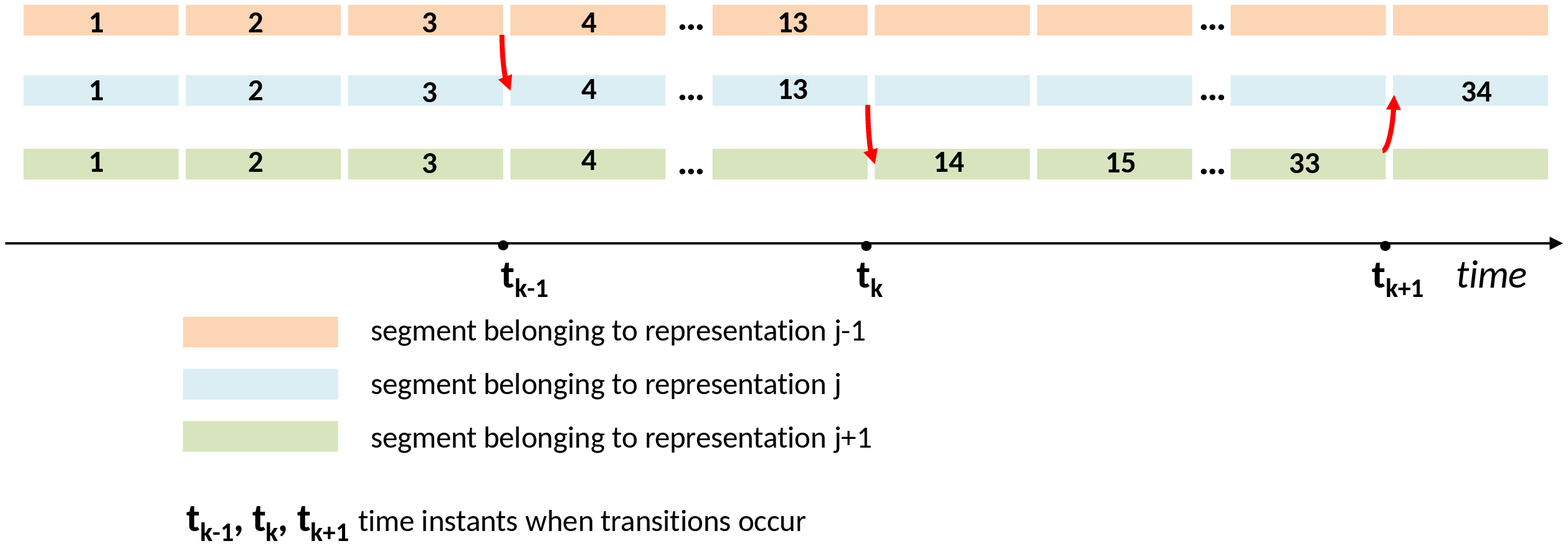}
\caption{Migrating to different media representations in ideal conditions} 
\label{fig:chunk-duration}
\end{figure}

\begin{figure}[h]
\centering
\includegraphics[trim = 10mm 100mm 15mm 50mm, clip,width= \columnwidth]{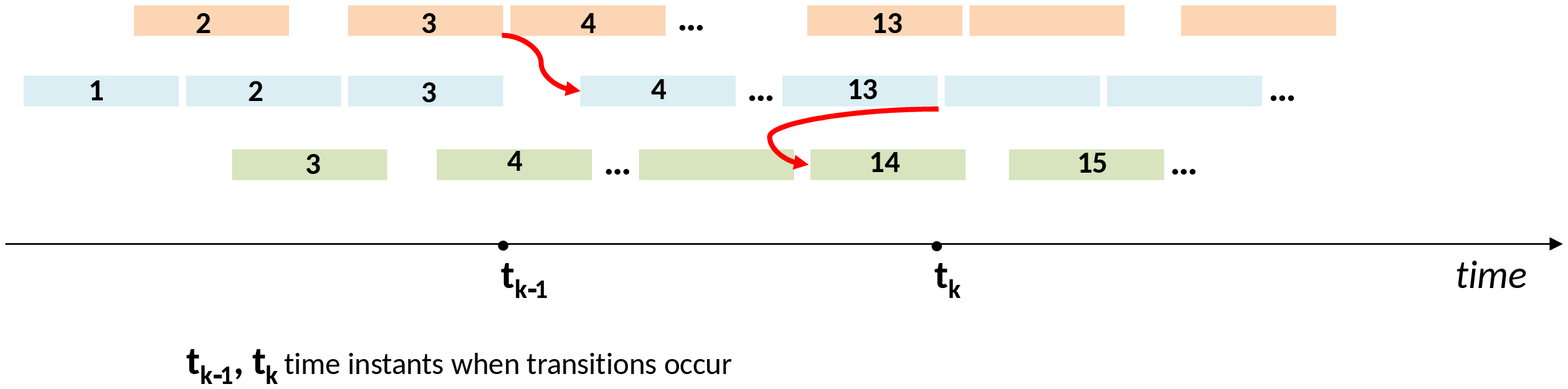}
\caption{Migrating in real conditions} 
\label{fig:more-realistic}
\end{figure}

A peer requesting the video channel for the first time begins watching representation $1$, 
that features the lowest bit rate, $r_1$, hence the lowest quality:
this choice allows to start the playout in a reasonably short time, and positively influences 
the overall quality  that the user experiences \cite{ZAMBELLI}. 
After joining overlay 1, if the peer's desired representation is not the video alternative at the lowest rate, the peer
attempts to move upward and possibly succeeds; however,
during the flow of the streaming the peer might also have to move downward to a lower quality representation,
depending on  current system  conditions, 
and then it will dynamically attempt to advance again. 
These actions translate into migrations from one overlay
to another and we require that a peer exclusively moves from its current overlay to an adjacent one,
i.e.,  from overlay 
$j$ to $j+1$ or $j-1$ if $2\leq j \leq K-1$,
from overlay $j$ to overlay $j+1$ if $K=1$, 
and from $K$ to $K-1$ if $j=K$; 
this guarantees the minimal gap between two consecutive representation playouts,
as recommended  in  \cite{QOEDASH}\cite {QOELAYERED} to confine  the amplitude of bit rate variations and its
negative effects on the perceived video quality. 
Furthermore, we assume that every DASH segment is made
of the same number $n$ of video chunks and that the chunk duration $t_{chunk}$ is the same in every overlay, which translates 
into a different amount of media bits being placed in every chunk.
Indicating by $L_j$ such size for the chunks that are distributed within overlay $j$, we have
${L_j}=t_{chunk}\cdot{r_j}$;
as Fig.~\ref{fig:chunk-duration} shows for the ideal case of  perfectly synchronized overlays,
this choice allows a smooth transition between different representations.
In the realistic circumstance of a time lag between different overlays as 
portrayed in Fig.~\ref{fig:more-realistic},
the constant duration of the chunks still preserves the smoothness of the transitions, as long as the user's buffer can absorb time shifts among different representations. 

\subsection
{The Proposed Algorithm}
In a conventional client-server scenario, 
DASH provides the user with the functionalities needed to perform adaptive streaming and leaves open the 
selection of the algorithm that rules the switching among the different available representations. 
The client is assigned the task to check its current conditions via the monitoring of
various parameters, and to ask the server for a different representation, if needed.
As previously observed, proposals in literature rely on the observation of several indicators and
put forth different criteria to govern the switching.
However, it is exclusively the local perspective that plays a role in the design of the control algorithm:
within the current proposal, although
the primary aim of the peer remains to experience a satisfying
streaming quality and to efficiently deploy its network resources, the user's decisions cannot any longer be taken in isolation, as
the peer has to be fair to all other peers within the system.
We therefore require that every peer employs two distinct types of status indicators,
\emph{local} and \emph{global}, to decide if, when and where to migrate; whereas the former indicators indirectly
supply information about the peer's video quality and are locally measured by the peer itself,
the latter parameters  provide a clue about the current health of the overlays and are periodically
handed out by the server to all peers, that rely upon both to enforce the adaptive rate selection algorithm. 

Among local indicators, we utilize:
\begin{itemize}
\item the Delivery Ratio (DR), defined as the ratio 
between the number of video chunks that
meet the playback deadline over the total number of chunks that a peer
should receive, measured with a periodicity of $\tilde t$ s.
Such ratio indicates the throughput that the peer experiences and
indirectly signals the quality of the received video in the 
very recent past; 

\item  the Request Window State (RWS), defined as the ratio between the number of downloaded video chunks 
within the current request window $W$ and the size of such window measured in number of video chunks, $0\le RWS \le 1$.
This indicator provides an indirect forecast of  video quality in the near future, as its value reflects the 
imminent status
of the playout buffer at the peer's site. 
\end{itemize}

Among global indicators, we select those among the status pointers 
determined by the  server that can be 
distributed to peers with very little effort, namely:
\begin{itemize}
\item the instantaneous resource index \cite{VUD} of the $j$-th overlay, $\sigma_j(t)$, $j=1,2,\ldots, K$, defined as
\begin{equation}
\sigma_j (t)= 
\frac
{C_{Sj} + \sum_{i \in N_j(t)} c_i}
{|N_j(t)|\cdot r_j}
\, ,
\end{equation}
where $C_{Sj}$ is the capacity that the  server commits to the $j$-th overlay to distribute
the representation with  rate $r_j$, $c_i$ is the upload capacity of the $i$-th peer belonging to overlay $j$, 
$N_j(t)$ is the set of active nodes   
within such overlay at time $t$ and $|N_j(t)|$ is the set cardinality; 

\item the instantaneous efficiency of the j-th overlay, defined as 
\begin{equation}
E_j(t)= \frac{U_{Sj}(t) +  	\sum_{i \in N_j(t)} u_i(t)}{|N_j(t)|\cdot r_j}
\label{eq:efficiency}
\end{equation}
where $U_{Sj}(t)$ is the actual upload rate that the  server provides to overlay $j$ at time $t$
and $u_i(t)$ is the actual upload rate at time $t$ of the $i$-th user within the same overlay.
\end{itemize}
When the instantaneous resource index $\sigma_j(t)$ takes on a value greater than or equal to $1$, in principle the $j$-th overlay 
can  successfully
guarantee the video delivery to all of its members, whereas when its value falls below $1$, the overlay operates in 
a critical regime; hence, $\sigma_j$ is a high-level indicator of the overlay health.  
However, the instantaneous efficiency $E_j(t)$ provides a more accurate picture, as
it captures some system behavior that would otherwise go unseen, if $\sigma_j$ only were examined.
To depict a scenario where this happens, it suffices to examine a flash crowd of viewers that
abruptly enters the system, wishing to stream the video: newly incoming peers initially
act as free riders, as they momentarily have no video chunks to share.
It is immediate to conclude that the value of 
the instantaneous efficiency drops, revealing a potentially critical operating condition, 
that the resource index cannot seize.

Let us next consider the generic peer $i$ within overlay $j$ and indicate by $r_d(i)$
the streaming rate of
its desired representation; the steps that the algorithm enforced by the peer goes through
every $\Delta t$ seconds are listed below: 
\begin{itemize}
\item[(i)] peer $i$ checks its current streaming rate $r_j$ against $r_d(i)$ and if $r_j<r_d(i)$, i.e., if the peer is not
satisfied, it first verifies whether it can leave its current overlay or it has to defer its departure.
This last circumstance occurs  
if overlay $j$ is not in good health and node $i$ upload capacity 
is beneficial to it, that is, if
$\sigma_{j}$ is lower than $1$ and if the peer upload capacity $c_i$ is greater than the streaming rate $r_{j}$; 
in this case, it is convenient that the peer does not move upward to overlay $j+1$. 
If on the contrary nothing prevents the departure, peer $i$ further verifies if 
its future contribution to the target overlay $j+1$ will be positive, i.e., if 
its upload capacity $c_i$ is greater than the streaming rate $r_{j+1}$.
If so, the peer migrates, 
as it will be beneficial to overlay $j+1$.
If not, the node further verifies that 
overlay $j+1$ has abundant overall upload capacity and 
that the video diffusion process within the overlay  is taking place in an efficient and noncritical manner, hence
the target overlay is able to accommodate a new peer, regardless of it being a relatively ``poor'' contributor.
So, the peer moves to overlay $j+1$ if 
$\sigma_{j+1}>1$ {\em{and}}  $E_{j+1}(t)> E_{thres}$, where  $E_{thres}$ is a properly set threshold.
 
\item[(ii)] peer $i$ also verifies its local status quality indicators, 
updating the 
weighted moving average of its delivery ratio and request window state in the following manner:
\begin{equation}
DR_i^{(t)}=w_D \cdot DR_i^{(t)}  + (1-w_D) \cdot DR_i^{(t-\Delta t) }
\label{eq:DR_estimate}
\end{equation}
and
\begin{equation}
RWS_i^{(t)}=w_W \cdot RWS_i^{(t)}  + (1-w_W) \cdot RWS_i^{(t-\Delta t) }
\, ,
\label{eq:RWS_estimate}
\end{equation}
respectively, where $w_D$ and $w_W$ are the weight coefficients. 
If $DR_i^{(t)}$ and also $RWS_i^{(t)}$ are below their predefined thresholds, $RWS_{thres}$ and $DR_{thres}$ respectively, 
the viewing quality is not deemed satisfying and the peer  scales down to a lower rate representation, hence it moves to overlay $j-1$.
\end{itemize}

The pseudo code describing the algorithm locally implemented by the peer is reported in Algorithm \ref{alg:SCA}. 

\begin{algorithm}[t]
\caption{ Rate Switching Control Algorithm}
\label{alg1}
\begin{algorithmic}
\STATE Node $i$ in overlay $j$ every $\Delta t$ seconds
\STATE \emph{;verifies its satisfaction}
\IF{( $r_j < r_{d}(i)$)}
\STATE\emph{;verifies the current overlay status}
\IF{( ($\sigma_{j}<1$) \AND ( $c_{i}>=r_{j}$))}
\STATE do not  migrate to  overlay ${j+1}$;
\ELSE
\STATE\emph{;verifies the destination overlay status}
\IF{( ( $c_{i}>r_{j+1}$) \OR
  ($\sigma_{j+1}>1$ \AND   $E_{j+1}>E_{thres}$) )}
\STATE migrate to overlay ${j+1}$;  exit;
\ENDIF
\ENDIF
\ENDIF

\STATE \emph{;verifies its viewing quality}
\IF{( ( $DR_{i}^{(t)}<DR_{thres}$) \AND ($RWS_{i}^{(t)}<RWS_{thres}$) )}
\STATE migrate to overlay ${j-1}$; exit;
\ELSE 
\STATE stay in overlay $j$; exit;
\ENDIF
\end{algorithmic}
\label{alg:SCA}
\end{algorithm}

Note that in the rate selection algorithm that is proposed to steer the peer's movements among overlays, 
the upgrade decision is conservatively influenced by the peer's local metrics 
-- its current rate $r_j$ and its desired streaming rate $r_d(i)$ -- and also by the global indicators of
the current and destination overlays, whereas 
the downgrade process is exclusively governed by local indicators.
The rationale behind this choice is that 
global indicators point to the overall status of the system 
and therefore let
the peer learn if the migration from its current swarm to the overlay distributing a higher bit rate representation 
can be safely performed. 
On the other hand, through local  indicators  the peer indirectly measures 
its streaming quality 
and decides if it is time 
to scale down to a lower DASH representation, in order to preserve a good viewing experience.
\section{Integer Linear Programming Model}
\label{sec:ILP}
To demonstrate that the proposed architecture and the rate control rule that it adopts
are able to cope with the expectations of the users in terms of desired streaming rate, 
we formulate an Integer Linear Programming (ILP) problem modeling a simplified, static system that will represent an initial  term 
of comparison.
More accurately, we examine $K$ overlays/representations and a population of $N$ users with heterogeneous streaming rate requirements,
and state that the system goal is to satisfy as many users as possible providing them with
the streaming rate they desire, subject to the only constraint that in every overlay the resource index takes on a value greater than one.
To mathematically state such problem,
we  
define user $i$ satisfaction within overlay $j$ as:
\begin{equation}
s(i,j) = 
\left\{
\begin{array}{ll}
1 \, , & \mbox{if $r_d(i)=r_j$} \\
0 \, , & \mbox{otherwise}
\end{array}
\right.
\, 
\label{eq:user-satisfaction}
\end{equation}
and state that the goal is to maximize the satisfaction function $\bar{S}$, defined as 
\begin{equation}
\bar{S}=\sum_{i=1}^N \sum_{j=1}^K s(i,j) \cdot x_{ij} 
\, ,
\label{eq:global-satisfaction}
\end{equation}
subject to the following constraints:
\begin{equation}
\sum_{j=1}^K x_{ij}=1 \, , \quad \forall i, \, i=1,2,\ldots, N
\, ,
\label{eq:x_constraint}
\end{equation}

\begin{equation}
r_d(i) - r_j \geq 0 \, , \quad if\quad  x_{ij}=1, \quad \forall i, \, i=1,2,\ldots, N
\, ,
\label{eq:rate_constraint}
\end{equation}
and
\begin{equation}
C_{Sj} + \sum_{i=1}^{N} c_i \cdot x_{ij}  - r_j\cdot \sum_{i=1}^{N} x_{ij}  \ge 0
\, , \quad \forall j, \, j=1, 2, \ldots, K
\label{eq:sigma_constraint}
\end{equation}
where 
$x_{ij}$ is a binary variable that indicates if peer $i$ is assigned to overlay $j$, i.e., 
\begin{equation}
x_{ij} = 
\left\{
\begin{array}{ll}
1 \, , & \mbox{if peer $i$ belongs to overlay $j$} \\
0 \, , & \mbox{otherwise}
\end{array}
\right.
 \, .
\end{equation}

The set of constraints in (\ref{eq:x_constraint}) guarantees that 
every peer belongs to one overlay only.
The set of constraints in (\ref{eq:rate_constraint}) indicates that
the $i$-th peer can only belong to an overlay that distributes the video at a rate 
$r_j$ lower than or equal to its desired rate $r_d(i)$. 
The set of constraints in (\ref{eq:sigma_constraint}) is equivalent to the requirement that in every overlay
the resource index $\sigma_j$ is greater than $1$.
To verify last statement, it suffices to observe that 
$r_j\cdot\sum_{i=1}^{N}  x_{ij}=r_j \cdot|N_j| $: dividing both members of
(\ref{eq:sigma_constraint}) by it easily demonstrates the equivalence.

The solution to this problem answers the following question: 
if a genie were to allocate as many users as possible in the way they desire, exclusively respecting the resource index constraint
in every overlay, what is the maximum value of satisfaction that could be attained?
Note however that two important remarks apply: 
\begin{itemize}
\item the ILP model represents a simplified P2P-DASH system operating in a static setting, where no notion of 
quality is present and where peers do not dynamically enter and leave the system; as such, it provides an upper, optimistic bound to the actual global satisfaction that any
newly proposed system exhibits in a real scenario;
\item There might be more 
distributions of the peers within the overlays that all achieve the same, maximum value of satisfaction.
It is therefore useful to verify whether we have a unique
distribution or not: if the former circumstance occurs, we can consider the actual values it provides and compare against them the distribution determined by a newly proposed control.
\end{itemize}
In order to check whether we have a unique problem solution and consequently one single distribution for the online nodes in each overlay, we proceed in the following manner: we numerically solve the problem and  determine the achieved value of satisfaction
$S$,
then proceed to a new formulation, where we add to the previous constraints in (\ref{eq:x_constraint}) and 
(\ref{eq:sigma_constraint}) the following:
\begin{equation}
\sum_{j=1}^K
\sum_{i=1}^{N}
s_{ij} \cdot x_{ij}
\ge  S
\, ,
\end{equation}
and replace the objective function in (\ref{eq:global-satisfaction}) with a new one, namely
\begin{equation}
S'=\sum_{i=1}^N \sum_{j=1}^K s(i,j) \cdot w_{ij} 
\, ,
\label{eq:new_global-satisfaction}
\end{equation}
the $w_{ij}$'s being uniformly randomly picked values in $[0,1]$.
If the distribution of peers within the different overlays that fulfills the new maximization problem  does not vary with respect to the original solution, then the 
distribution is unique and can be profitably
employed for comparison purposes.  
We anticipate here that applying this procedure to the examined setting revealed that the solution we determined is indeed unique. 
\section{Numerical Results}
\subsection{Simulation Setup}
\label{sec:Simulation Setup}
To evaluate the performance of the system under investigation  and of the algorithm that it employs, we implemented an event-driven 
simulator based on the source code  available in \cite{ZHANG}. We  
built a replica of a multi-overlay system, where every swarm 
streams the same video at a different bit rate.
The  system has an average population of $N = 2000$ active peers, that dynamically enter and leave; 
nodes populate the system within  the first $20$ s and in this interval their interarrival times 
are exponentially distributed with an average of $0.1$ s; after the first $20$ s, 
the interarrival times are modulated so as to keep nearly constant the number of peers in the system.
Session times are exponentially distributed,
with an average of $1500$ s.
Nodes belong to four different classes whose
upload and download capacity values stem from the current European Internet connection offerings \cite{SWISSCOM}\cite{ORANGE}\cite{TNET};
the percentages of users belonging to the different classes
are drawn from the Akamai European average connection speed report \cite{AKAMAI}; the employed values are reported
in Table \ref{tab:Users_profiles}. 
$K=4$ video representations are available at rates 
$700$, $1500$, $2500$ and $3500$ kbit/s;
such values have been chosen
having in mind the typical streaming rates of Internet Standard Definition video (SD) and High Definition video (HD).
The streaming server allocates to each overlay only a small amount of its upload capacity, equal to four times the rate
of the video, i.e., $C_{Sj}=4\cdot r_j$, indicating that our focus is on a pure P2P system.
Moreover, the size of the current request window that every peer works with is $X=20$ s, the chunk duration is $t_{chunk}=200$ ms, the delivery ratio is locally computed every $\tilde t= 5$ s
and the threshold values we employed are the following: $DR_{thres}=0.5$, $RW_{thres}=0.3$ and $E_{thres}=0.9$. In next subsections we will provide a thorough justification for these choices.
The coefficients for the computation of the weighted moving average of
$DR_i$ and $RWS_i$ are $w_D=\frac{1}{3}$ and $w_W=\frac{2}{3}$, respectively,
revealing that, with regard to the local delivery ratio we tend to privilege
stability in the estimate in (\ref{eq:DR_estimate}), whereas for the request window state in (\ref{eq:RWS_estimate}) 
we give priority to what happens at current time.
Last, the periodicity that the rate control algorithm employs is $\Delta t=4$ s.

\begin{table}[htbp]
\centering
\begin{tabular}{|l|r|r|r|r|}
  \hline
   & \emph{Class 1} & \emph{Class 2} & \emph{Class 3} & \emph{Class 4}\\
  \hline
\emph{Upload capacity  (kbit/s)}  &  704 & 1024&1500&10000 \\
\hline
\emph{Download capacity (kbit/s)} & 2048 & 8192& 10000&50000\\
\hline
$\%$ of peers& 20  &  21& 42&17\\
  \hline
\end{tabular}
\vspace{5mm} 
\caption{User capacity profiles and percentages} 
\label{tab:Users_profiles}
\end{table}
We begin investigating system behavior in three indicative scenarios, termed
conservative, uniform and aggressive: the aim is to quantify system behavior in distinct, meaningful settings, each
representing a reference case that might indeed occur.
Within the conservative setting, every peer is allowed to prudently stream the highest representation whose bit rate 
is lower than its upload capacity: so, class $1$ and class $2$ users can only stream representation $1$, 
class $3$ users representation $2$ and class $4$ users representation $4$. 
Within this scenario every overlay is exclusively composed by peers 
whose upload capacity is greater than the streaming rate of the video that the overlay distributes and bandwidth resources are therefore 
abundant.
The second investigated scenario is termed uniform, as the generic peer uniformly and randomly selects the
representation it desires,
subject to the natural constraint that it is lower than the peer's download capacity.
The underlying idea is to mimic the spreading of different representation requests 
due to the users' displays, with heterogeneous resolution capabilities.
It follows that the representation range is limited  for class $1$ users only, that choose 
between representation $1$ and $2$ with equal probability, whereas users belonging to class $2$, $3$ and $4$ equally distribute their 
video requests among the four available representations.
The third scenario is termed aggressive, as every peer aims at streaming the video at the highest representation whose bit rate 
is lower than its download capacity:  
in this case all peers  aim at streaming representation $4$, except for class $1$ peers, that are confined to 
representation $2$. 
In the aggressive scenario it is impossible to fulfill all users' requests: if peers were uncritically placed within the 
overlays distributing the desired representations ($2$ and $4$), 
the aggregate upload capacity of each swarm 
would not be enough to satisfyingly deliver all peers the video, the resource index values being markedly lower than $1$,
namely, $\sigma_2=0.48$ and $\sigma_4=0.91$. 
So, in this setting the challenge is to place as many peers as possible within the desired overlay, in a manner that 
secures good values to the local quality indicators.

The results of next subsection  numerically 
illustrate our findings.
\subsection{Evaluating Performance in Different Scenarios}

\begin{figure}[t]
 \centering
\includegraphics[width= \columnwidth]{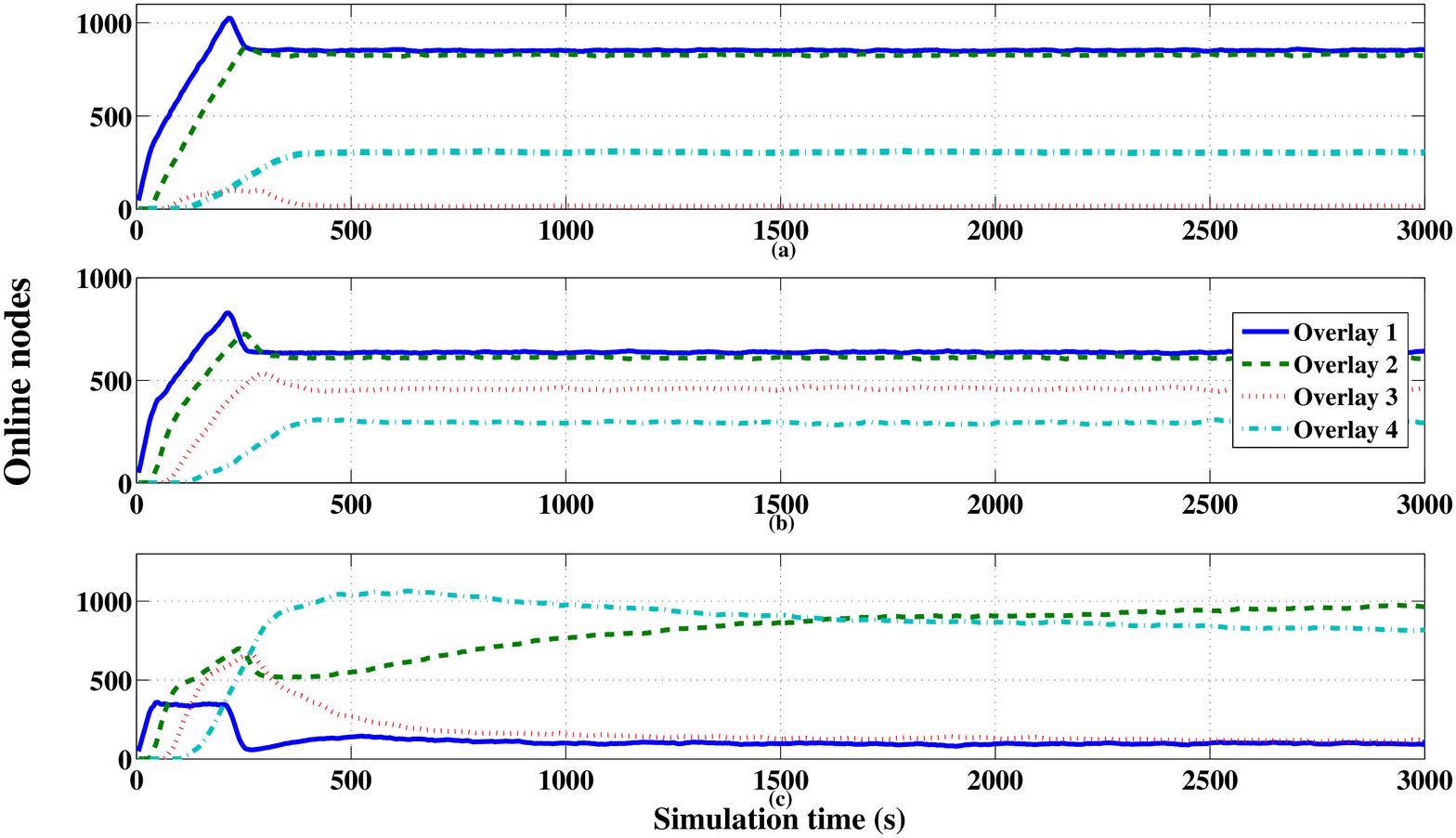}
\caption{Node distribution in the examined scenarios: (a) conservative, (b) uniform, (c) aggressive}
\label{fig:online-distribution}
\end{figure}

Fig.\ref{fig:online-distribution} reports the distribution of online nodes among the four overlays, each corresponding to one
video alternative, as a function of time.
Fig.\ref{fig:online-distribution}(a) refers to the conservative setting: after the initial transient,
when we compare the  number of nodes within each overlay shown in this figure to the number of nodes that wish  to stream every 
representation, reported in the first row of Table \ref{tab:scenarios}, we conclude that they are very close. 
On the contrary, Fig.\ref{fig:online-distribution}(b) shows that in the uniform scenario, the 
nodes that stream representation $4$ are fewer than the ones that would like to stream it, as indicated
in the second row of Table \ref{tab:scenarios}, revealing that it is not possible to serve all peers requesting the highest rate representation,
owing to the peers' scarce upload capacity. 
As a result, the rate control algorithm redirects some of these peers 
towards the overlays that distribute lower rate videos: this also 
explains why the number of online nodes watching representations  $1$ and $3$ slightly deviates from 
the users requests indicated in the same table row.
Fig.\ref{fig:online-distribution}(c) refers to the aggressive scenario: here, it is worth noting that the initial transient prolongs longer 
and that the number of nodes in each distinct overlay 
significantly differs from the number of peers wishing to stream the different representations indicated 
in Table~\ref{tab:scenarios}, third row.  
This happens because there are not enough resources to satisfy all users' requests, 
only a fraction of  nodes are placed in the desired overlay and the remaining ones must be redistributed among the other overlays.
Moreover, the unsatisfied nodes will try, whenever possible, to move to a higher rate representation.

\begin{figure}[t]
 \centering
\includegraphics[width= \columnwidth]{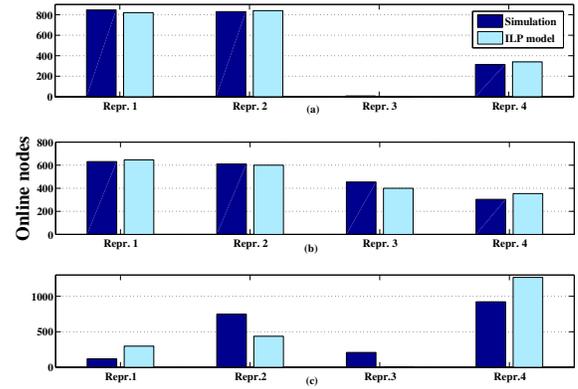}
\caption{Comparison between the P2P-DASH architecture and the ILP model in the examined scenarios: (a) conservative, (b) uniform, (c) aggressive }
\label{fig:SIM-vs-ILP-distribution}
\end{figure}

Fig. \ref{fig:SIM-vs-ILP-distribution} illustrates the comparison between the number of online nodes 
within each overlay obtained by simulation and the number provided by the numerical solution of the ILP problem, for the three examined scenarios. 
The number of online nodes within the simulated P2P-DASH system has been averaged over time  
(considering only the last $1500$ s of the simulation) 
and also over $10$ distinct simulation runs. 
In the conservative and uniform scenarios,  Figs.\ref{fig:SIM-vs-ILP-distribution}(a)-(b) indicate that
the two distributions are extremely close.
This proves that our algorithm correctly controls the peer movements when system resources are not overly stressed.
In the aggressive scenario, Fig.\ref{fig:SIM-vs-ILP-distribution}(c)
shows that the difference is more noticeable.

\begin{figure}[htbp]
 \centering
 \includegraphics[width=\columnwidth]{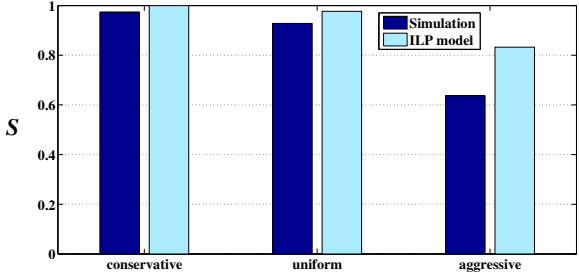}
\caption{System satisfaction normalized to the number of users}
\label{fig:sat-comp}
\end{figure}

\begin{table}[htbp]
\begin{center}
\begin{tabular}{|c|c|c|c|c|c|}
  \hline
  \emph{Scenario} & \emph{repr. 1} & \emph{repr. 2}& \emph{repr. 3}& \emph{repr. 4}  \\
  \hline
\emph{conservative} & 820 & 840 &  - &340 \\
\hline
\emph{uniform} & 600 & 600& 400&400\\
\hline
\emph{aggressive} &  - &  400& - &1600\\
  \hline
\end{tabular}
\vspace{5mm} 
\caption{Number of nodes wishing to stream each representation} \label{tab:scenarios}
\end{center}
\end{table}
Fig.\ref{fig:sat-comp} further shows the comparison between the  system satisfaction $S$ that the proposed P2P-DASH architecture 
attains and the satisfaction value
provided by the  numerical solution of the ILP problem for the reference system of Section \ref{sec:ILP}; the reported values have been normalized with respect to 
$N=2000$, the total number of users. Within the P2P environment,
satisfaction  has been monitored during every simulation run with a periodicity of $10$ s over the last $1500$ s  
of the streaming session, then these instantaneous values have been averaged over time and also over $10$ distinct simulation runs.
The figure shows that our system nearly satisfies as many nodes as the system modeled through the ILP approach in both
the conservative and the uniform scenarios; this further corroborates  the conclusion that the proposed solution is effective at exploiting  resources when no critical shortage occurs.
It does not, only when the most demanding, aggressive scenario is examined: however,
as next figures will demonstrate, the proposed system
succeeds at guaranteeing its users  good  local conditions, 
a goal the ILP model cannot pursue.

\subsection{Tuning the Rate Control Parameters }
\begin{figure}[t]
 \centering
 \includegraphics[width=\columnwidth]{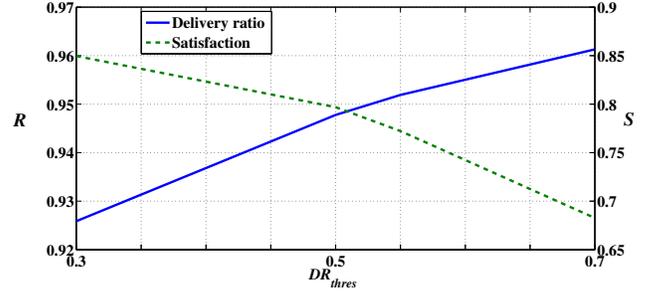}
\caption{Satisfaction and delivery ratio as a function of $DR_{thres}$ }
\label{fig:DR-thr-effect}
\end{figure}

From now onward, we will exclusively concentrate on the
most challenging setting, which is the aggressive scenario, and we will
begin tuning the thresholds that the rate control algorithm employs so as to optimize system performance.
We first investigate the $DR_{thres}$ effect on  system behavior, and take as
reference metrics the delivery ratio that the overlays exhibit, averaged  over their population,
and the system satisfaction. 
In order to have the downward migration process exclusively ruled 
by the comparison between $DR_i^{(t)}$, the value of the peer local delivery ratio,  and  the
$DR_{thres}$ value, 
we set $RW_{thres}=1$ in Algorithm \ref{alg:SCA}.
Fig.\ref{fig:DR-thr-effect} accordingly shows the satisfaction $S$ and the delivery ratio $R$ that the system achieves, when varying the $DR_{thres}$ threshold.
The delivery ratio $R$ that appears in the figure has been computed as the weighted average of the average delivery ratio of each overlay, where the weight associated to each value corresponds to the average number of users within the same overlay. 
We note that $R$ increases as $DR_{thres}$ increases: as a matter of fact,
a greater value of the threshold translates into a looser constraint ruling the downward migration process. 
In other words, 
in every overlay other than the first a larger number of peers migrates to the lower overlay, while those who remain exhibit relatively high values of delivery ratio. 
On the other hand, being easier to move downward reflects into a lower chance for the peer to reside within the desired overlay, where-from a lower value of satisfaction $S$.
It is therefore necessary to find a compromise value for $DR_{thres}$, that at the same time
guarantees good values for the considered metrics. 
As Fig.\ref{fig:DR-thr-effect} suggests,
we chose to work with $DR_{thres}=0.5$,
a trade-off value that warrants a satisfying value for both.

  \begin{figure}[t]
 \centering
 \subfigure[$DR_{thres}=0.5$ $RWS_{thres}=1$ ]
   {\includegraphics[width=8.5truecm]{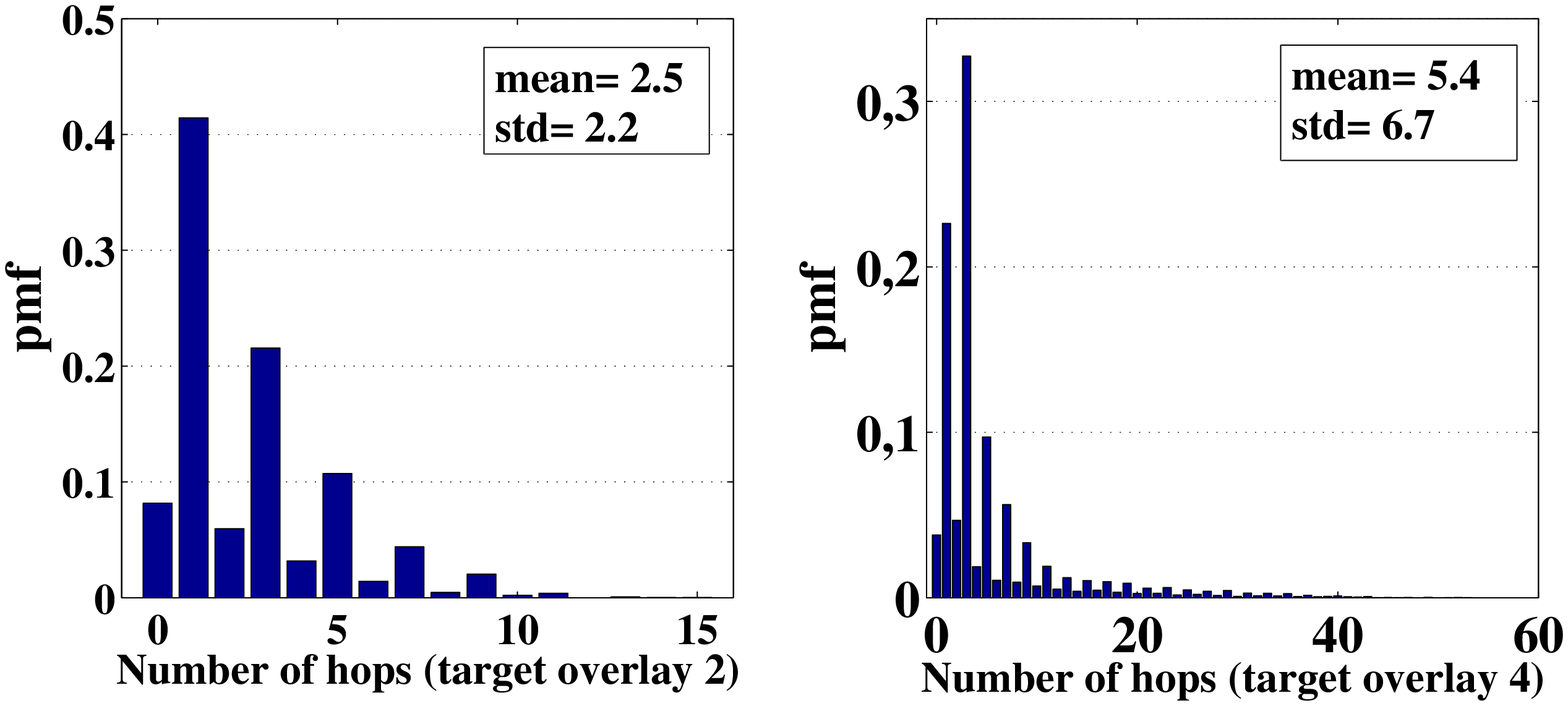}}
 \hspace{1mm}
 \subfigure[$DR_{thres}=0.5$  $RWS_{thres} =0.7$ ]
   {\includegraphics[width=8.5truecm]{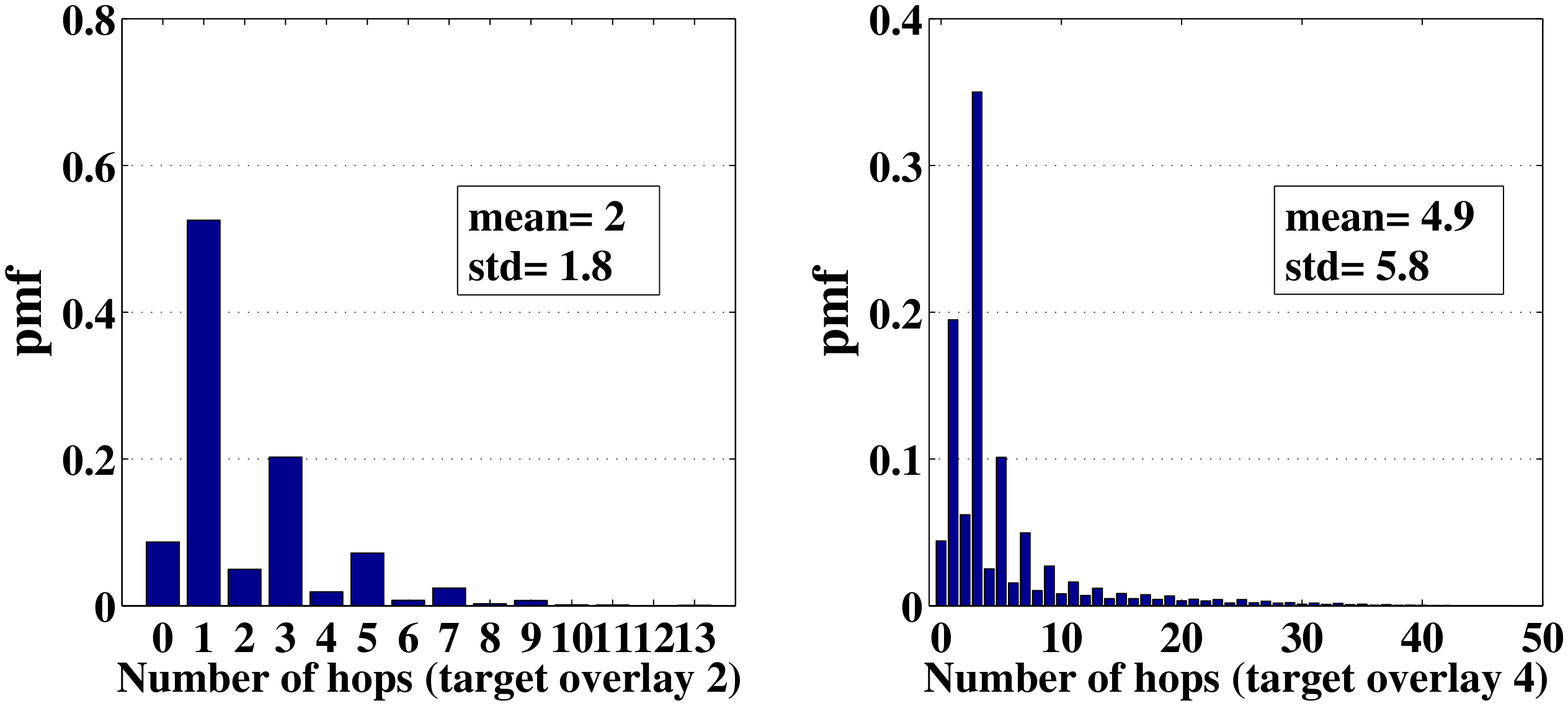}}
   \hspace{1mm}
 \subfigure[$DR_{thres}=0.5$  $RWS_{thres} =0.3$ ]
   {\includegraphics[width=8.5truecm]{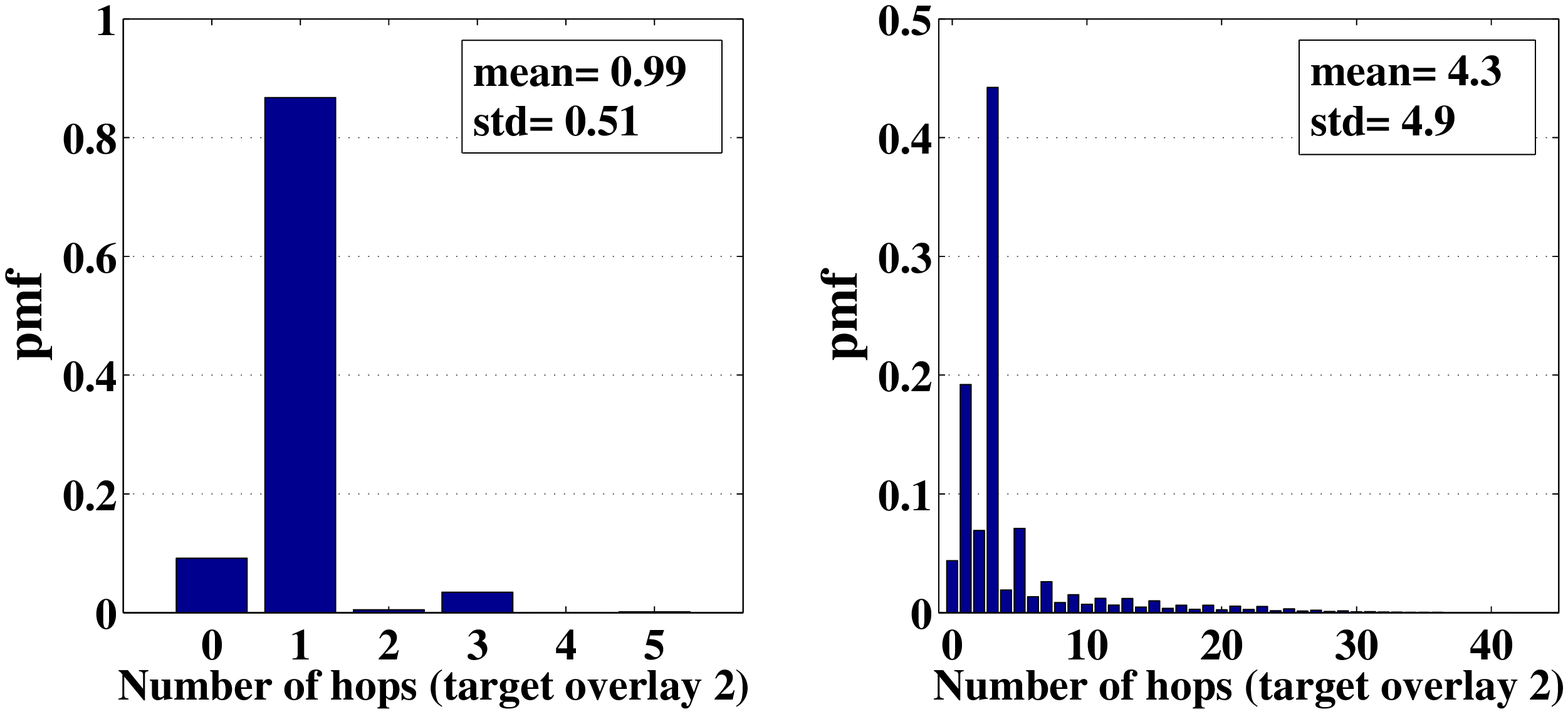}}
 \caption{pmf of the number of hops for peers targeting overlay $2$ and overlay $4$}
 \label{fig:RWS-switch-pdf}
 \end{figure}
  \begin{figure}[htbp]
 \centering
 \subfigure[$DR_{thres}=0.5$  $RWS_{thres}=1$ ]
   {\includegraphics[width=8.5truecm]{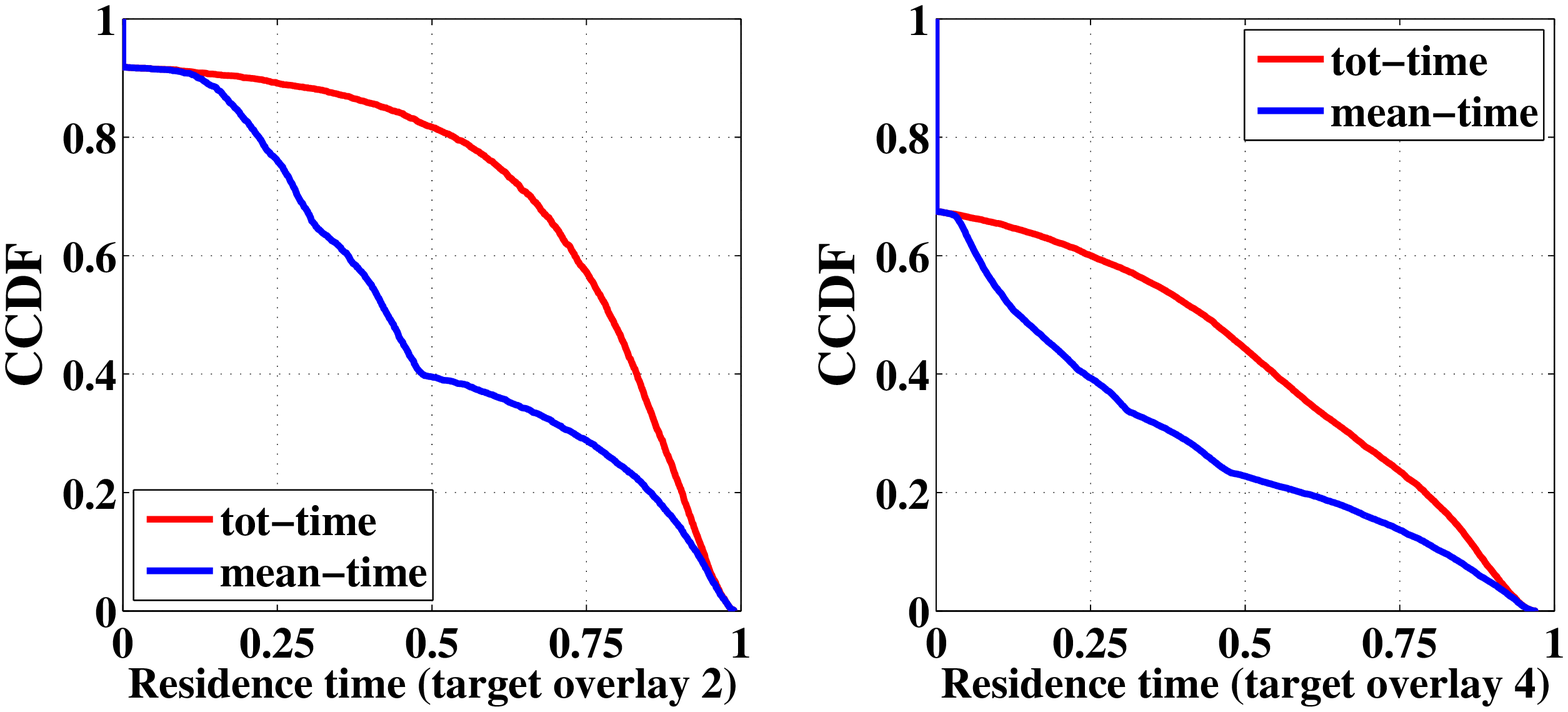}}
 \hspace{1mm}
 \subfigure[$DR_{thres}=0.5$  $RWS_{thres} =0.7$ ]
   {\includegraphics[width=8.5truecm]{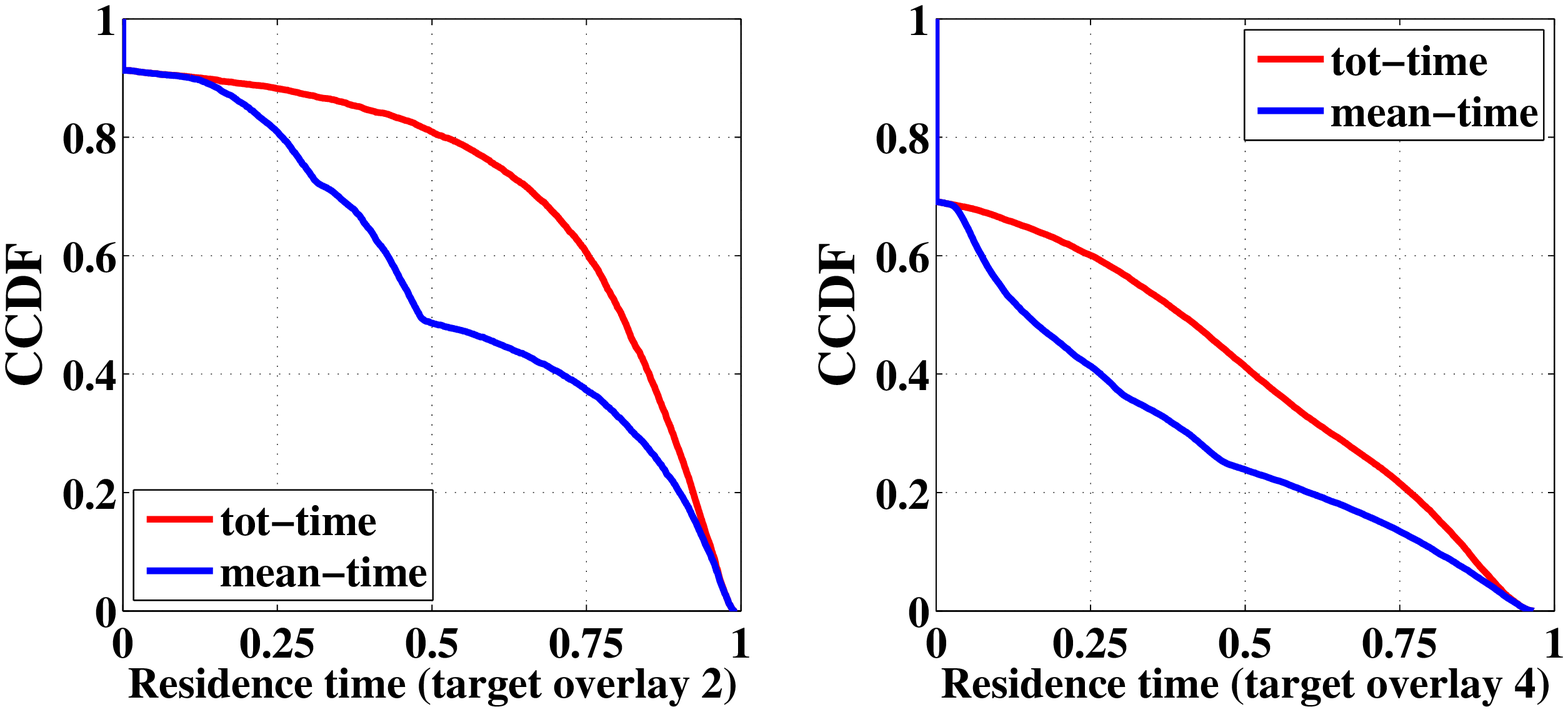}}
   \hspace{1mm}
 \subfigure[$DR_{thres}=0.5$  $RWS_{thres} =0.3$ ]
   {\includegraphics[width=8.5truecm]{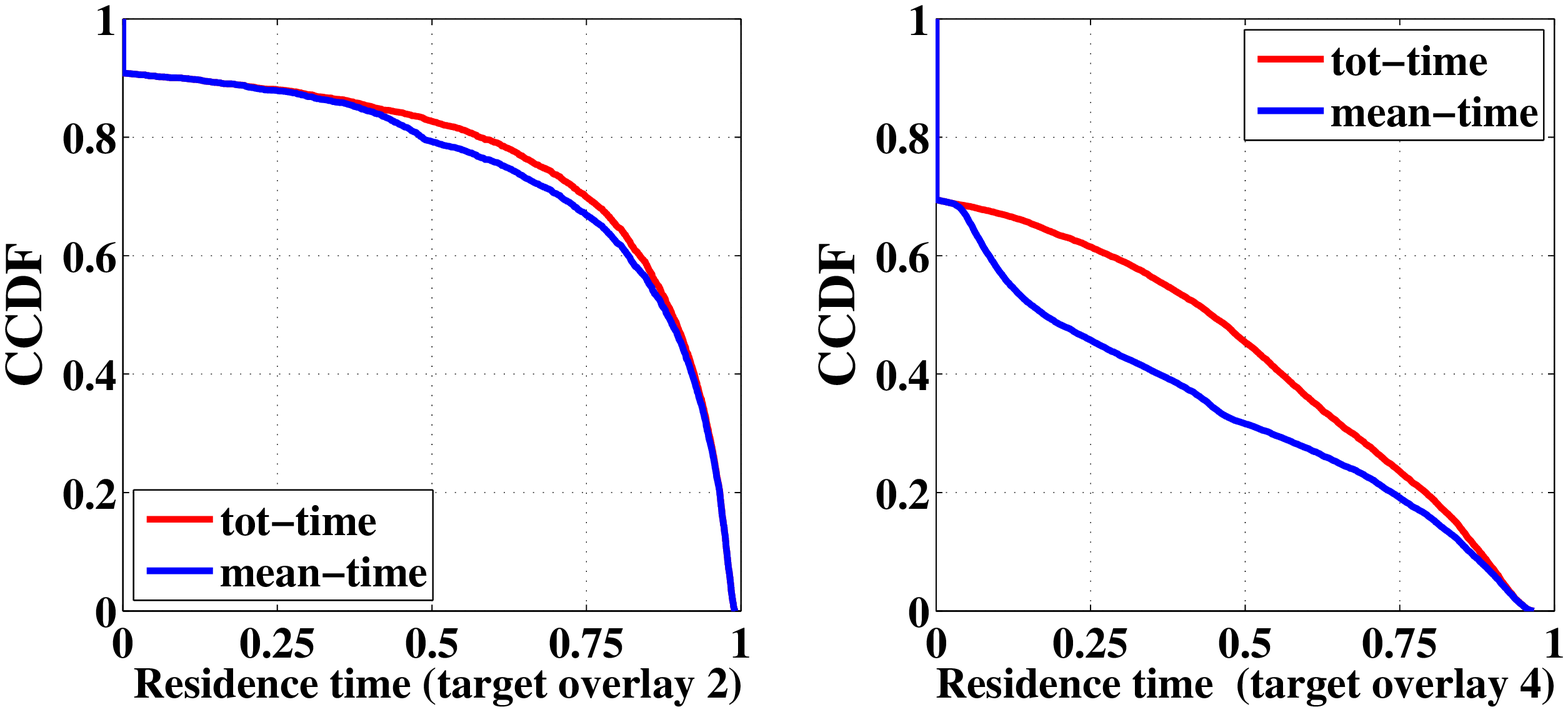}}
 \caption{CCDF of the normalized residence time}
 \label{fig:RWS-residence-time-ccdf}
 \end{figure}

We next examine the influence of $RW_{thres}$, the threshold on the request window state, on the algorithm behavior and in turn onto the achievable system performance. 
This threshold acts on the local parameter $RWS^{(t)}$, that provides 
the peer with a forecast of the status of its playout buffer in the very next future.
Whereas it is meaningful to consider the condition on the 
local delivery ratio alone, as we did before, we believe that the same does not hold for the condition
on the request window state: as a matter of fact, the former offers a reliable snapshot
of the current situation at the peer's site, the latter displays the intrinsic uncertainty of a projection, and this inevitably affects the reliability  of the rate control decision process.
On the other hand,
to consider both conditions when deciding whether to downgrade
can turn out beneficial, especially if  $RW_{thres}$ takes on a  sufficiently modest value. 
In order to validate this statement, we begin by observing that 
both the delivery ratio $R$ and the system satisfaction $S$ 
are not heavily influenced by different choices for $RW_{thres}$:
when decreasing $RW_{thres}$ from $1$ to $0.3$, we observed that they
modify from $0.95$ to $0.94$, and from $0.77$ to $0.76$, respectively.
What is significantly affected by this threshold is the number of hops and the  time
each peer spends within the overlay that distributes its desired streaming rate.
The number of hops is a crucial metric to monitor, as the viewing quality 
is heavily influenced by it \cite{QOEDASH}, \cite{QOELAYERED}.
We therefore fix $DR_{thres}=0.5$ and report in 
Fig.\ref{fig:RWS-switch-pdf} the probability mass function (pmf) of the number of hops that the generic peer
makes from one overlay to another during its lifetime for three reference values of  $RW_{thres}$, namely, $RW_{thres}=1$ (that corresponds to bypassing the condition),
$RW_{thres}=0.7$ and $RW_{thres}=0.3$. 
We built the pmf's on the basis of the target overlay that the peers aim at reaching, as a user targeting the overlay that distributes the highest representation will make the
largest number of hops with the highest probability. In the examined aggressive setting, the two target overlays are overlay $2$ and overlay $4$. The figures on the left column refer to peers that wish to stream representation $2$, the ones on the right column to peers that wish to stream representation $4$.
In detail, the comparison between
Fig. \ref{fig:RWS-switch-pdf}(a) and Figs. \ref{fig:RWS-switch-pdf}(b)-(c) indicates that the introduction of the $RWS $ condition reduces the number of the hops of nodes, the effect being substantial  for the nodes aiming at representation $2$; as an example, when $RW_{thres}=0.3$  a peer makes only one hop
with probability $0.87$.  
A similar advantage, although not so impressive, is experienced by peers requesting  representation $4$: for this class of peers the average number of hops reduces  from $5.4$ to $4.3$ and also the standard deviation decreases. 
  
Fig.\ref{fig:RWS-residence-time-ccdf} depicts the complementary cumulative distribution function (CCDF) of the total  time (red line) and the average consecutive time (blue line) the generic
peer spends within the desired overlay, normalized to the peer's life time. 
Fig.\ref{fig:RWS-residence-time-ccdf} (a) refers to the case where  no $RWS^{(t)}$ control is exerted, and Figs.\ref{fig:RWS-residence-time-ccdf} (b)-(c) refer to the adoption of the threshold $RW_{thres}=0.7$ and  $0.3$, respectively. 
We pleasingly observe that for the peers aiming at overlay $2$ the two CCDF's progressively get closer as $RW_{thres}$ decreases:
if $RW_{thres}$ is sufficiently low, the probability  that
the peer spends more consecutive time within the desired overlay significantly increases. 
An additional threshold, $E_{thres}$, comes into play when the peer locally implements
the rate control algorithm, but we deliberately postpone its investigation to
subsection \ref{ss:Flash-crowd}, where we analyze system behavior in the presence of a flash crowd.

\subsection{DASH-unaware System Comparison }
\begin{figure}[htb]
\centering
\includegraphics[width= \columnwidth]{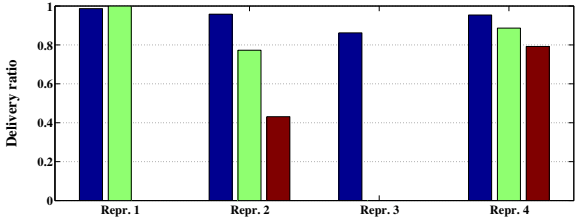}
\caption{Comparison among average delivery ratios: P2P-DASH architecture and systems with isolated swarms}
\label{fig:delivery-comparison}
\end{figure}

\begin{figure}[htb]
\centering
\includegraphics[width= \columnwidth]{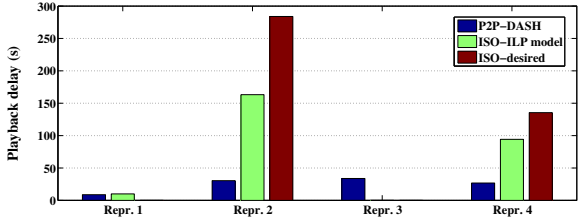}
\caption{Comparison among average playback delays: P2P-DASH architecture and systems with isolated swarms}
\label{fig:playback-comparison}
\end{figure}

Having tuned the algorithm, we next proceed to demonstrate the effectiveness of our proposal.
To this end,
Fig. \ref{fig:delivery-comparison} confronts the average delivery ratio 
that each overlay exhibits in the P2P-DASH architecture (blue bars) to the average delivery ratio of  
two DASH-unaware P2P systems made of 
isolated overlays. 
First, we consider a system where there is one overlay for every distinct streaming rate that peers might request. 
In this solution, nodes do not support DASH and therefore cannot dynamically move from one swarm to another;
rather, they join and reside in the overlay that distributes the video at the bit rate they request.
We term this solution as ``ISO-desired'' (red bars). 
In the third system we examine,
 peers  join and reside in the overlay that is indicated by the solution of the ILP problem  (see Fig. \ref{fig:SIM-vs-ILP-distribution}(c)), so as to obtain the optimal  value of global satisfaction; we refer to it as ``ISO-ILP'' (green bars).
The values of delivery ratio have been computed as averages over the total number of active peers and also over the streaming session duration,
for $10$ different simulation runs.
For the P2P-DASH system, although not all peers belong to the desired overlay, Fig.~\ref{fig:delivery-comparison} indicates that on average they
experience excellent values of delivery ratio,
always greater than $0.94$ except for the overlay distributing representation $3$, which is a transition overlay where 
very few nodes reside.
As for the ``ISO-desired'' system, only the two overlays distributing representation $2$ and $4$ are present and
both display unsatisfactory values of the delivery ratio
(as low as $0.43$ for overlay $2$,  about $0.79$ for overlay $4$).
Observing the behavior of
the ``ISO-ILP'' system, we can further outline  that except for overlay $1$, the average delivery ratio of overlays $2$ and $4$ is lower than  the one experienced by the corresponding overlays in the P2P-DASH system. 
As regards the playback delay, defined as the interval between the time
when the generic video chunk is generated at the video server and the instant
the same chunk is rendered at the peer's site, the improvement that the P2P-DASH architecture achieves is even more significant: Fig.\ref{fig:playback-comparison} reveals that the average playback
delay is approximately reduced by an order of magnitude.
Overall, the comparison definitely plays in favor of the proposed architecture.

\subsection{Flash-Crowd Event}
\label{ss:Flash-crowd}
Given dynamism is exactly the reason why adaptive streaming 
through DASH has been introduced, it is significant to explore how the examined system reacts to the presence of a flash crowd,
so common in a P2P environment. 
In this subsection we therefore investigate how the proposed architecture reacts to the occurrence of a step join of new users,
that massively want to enter the system in a relatively short amount of time. 
We therefore overload the system, which has reached the steady-state condition and accommodates $N=2000$ users, with $N^{'}=3000$ 
new incoming peers that begin entering the system at time $t_{FC}=3000$ s for the next $30$ s. The same percentages and capacity values that originally described the
composition of the peer population are applied to this bulk arrival of new users.

\begin{figure}[htbp]
 \centering
 \includegraphics[width= \columnwidth]{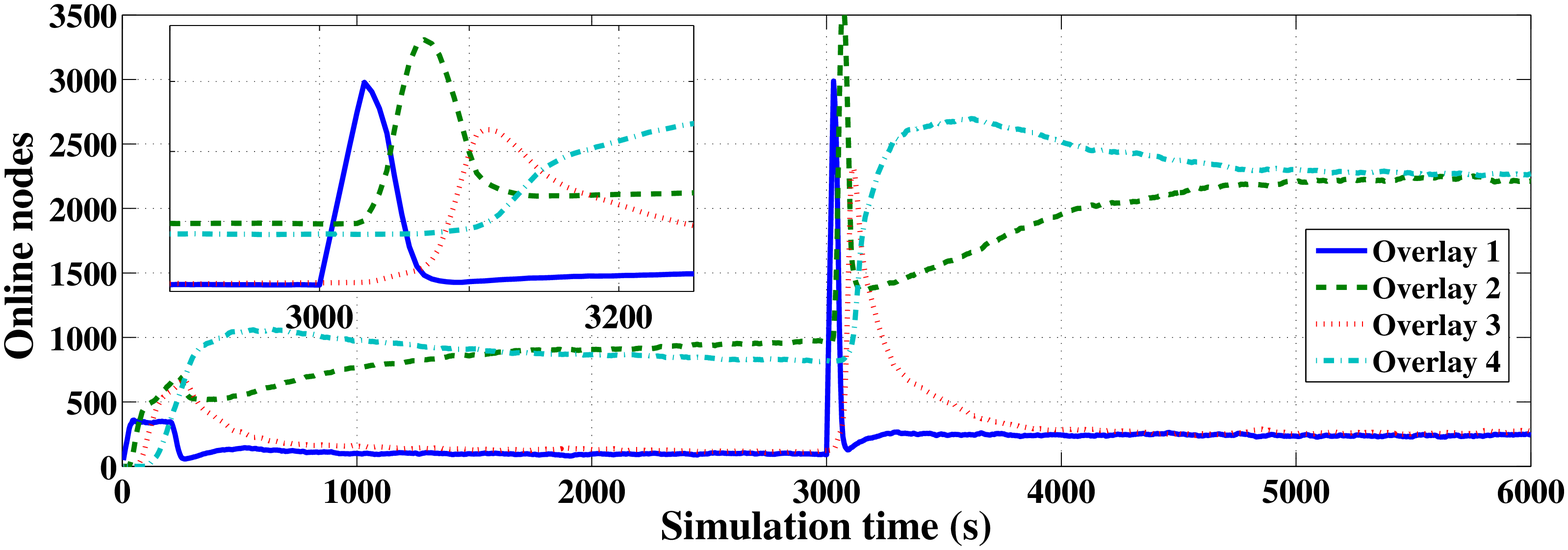}
\caption{Online nodes in the presence of a flash-crowd event}
\label{fig:online-FC}
\end{figure}

Fig.\ref{fig:online-FC} evidences the different tides of new peers moving from one overlay to the next: 
all users are forced to enter within the overlay that distributes the lowest representation of the video, and it is here that 
the number of online nodes first peaks; then they  move forward, their wave investing overlay $2$ and $3$, to finally reach overlay $4$. In less than $3$ 
minutes, the P2P platform is able to react to this massive strain.

\begin{figure}[htbp]
\centering
\includegraphics[width= \columnwidth]{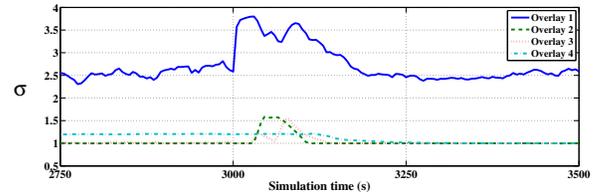}
\caption{Evolution of the resource index}
\label{fig:sigma}
\end{figure}

Within such framework, it is also crucial to discuss the relevance of the efficiency displayed by the system overlays as defined in (\ref{eq:efficiency}) and employed by the rate control algorithm as a global indicator.
To this end, we recall that in Algorithm \ref{alg:SCA} the upward migration of the peer is ruled 
by the combined observation of the current value of the resource index and of the efficiency that the target overlay displays.
The condition on the first indicator reflects the static resource balance, the condition on the latter points to the true capability that the target overlay owns to successfully deliver the video.
In greater detail,
Fig.\ref{fig:sigma} indicates that the massive arrival determines a huge peak in $\sigma_1(t)$, the resource index of the overlay distributing the lowest streaming rate, as all new peers enter overlay $1$ and all of them have an upload capacity $c_i$ greater than or equal to $r_1$; the same trend, although on a reduced scale, is observed for the remaining overlays, except for the one distributing the video at the highest streaming rate, 
where on the contrary the majority of the joining peers  have an upload capacity $c_i$ lower than $r_4$.

\begin{figure}[htbp]
\centering
\includegraphics[width= \columnwidth]{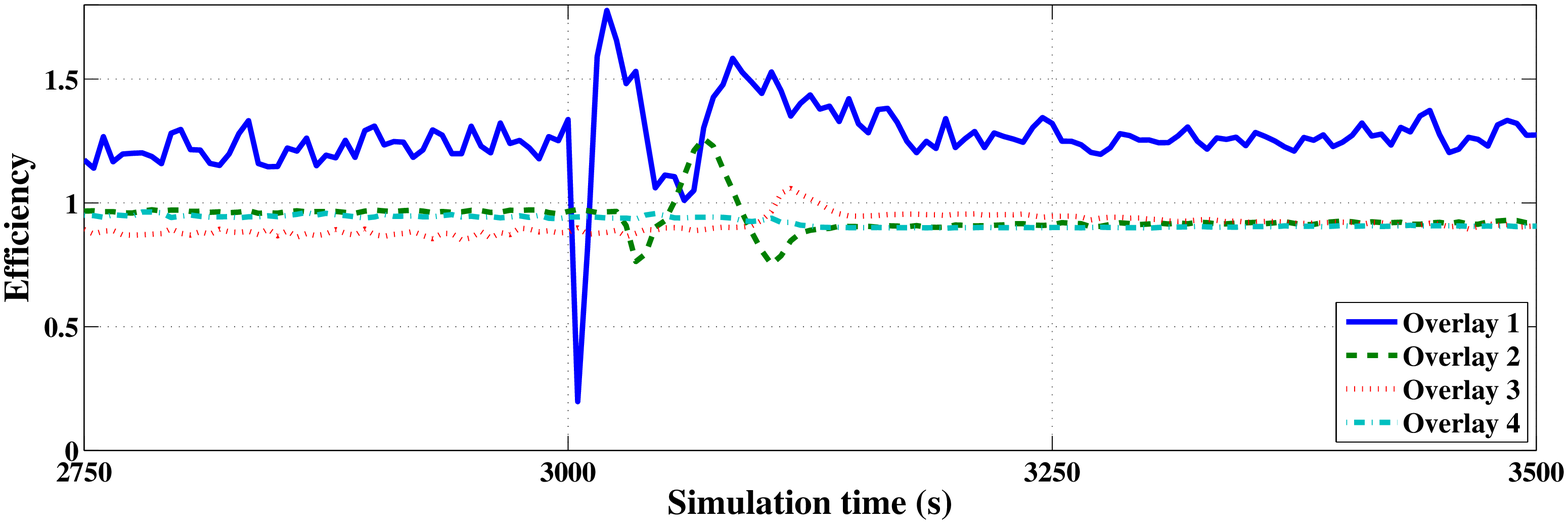}
\caption{Evolution of efficiency}
\label{fig:eff}
\end{figure} 
If in parallel we examine what happens to efficiency, we notice from Fig.\ref{fig:eff} 
that the flash crowd occurrence immediately determines a sharp, negative peak in $E_1(t)$, which
in turn reflects in $E_2(t)$: it is a manifest signal that a 
critical regime will soon appear. As a matter of fact, even a robust overlay can momentarily operate in critical conditions and the reason is that its newcomers will temporarily act as free riders: they do require video content, without sharing any.

\begin{figure}[htbp]
\centering
\includegraphics[width= \columnwidth]{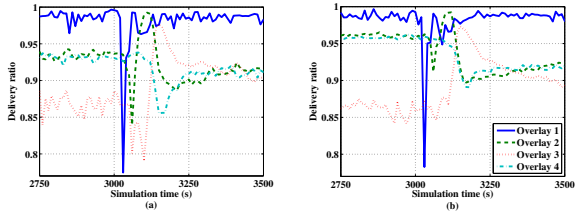}
\caption{Comparison between delivery ratios}
\label{fig:delivery-comp}
\end{figure}

Fig.\ref{fig:delivery-comp} shows the behavior of the average delivery ratio (the average being computed over the peer population in every overlay) when the flash crowd occurs, and compares the behavior of the system where the algorithm exerts no control on $E(t)$ (Fig.\ref{fig:delivery-comp}(a)) to the system where the algorithm duly takes it into account (Fig.\ref{fig:delivery-comp}(b)). We  set the threshold to the conservative value $E_{thres}=0.9$, so as to significantly slow down the upward migration during the abrupt step join. 
Fig.\ref{fig:delivery-comp}(b) indicates that the negative spikes of the delivery ratio  are either eliminated or markedly confined within all overlays except for overlay $1$. It is also interesting to notice that the average delivery ratio of  overlay $2$ and $4$ stays above $0.9$, indicating that the video diffusion process is taking place in a satisfying manner, and that in overlay $3$ the delivery ratio does not even show any reduction after $t=t_{FC}$. As for the negative impact that the flash crowd has on overlay $1$, it is unavoidable: this overlay represents the system entry point and as such the arrival of new peers is not influenced by the rate control algorithm.
\subsection{Buffer Content Reuse}
\begin{figure}[htbp]
\centering
\includegraphics[trim = 15mm 25mm 15mm 30mm, clip,width= 8.5truecm]{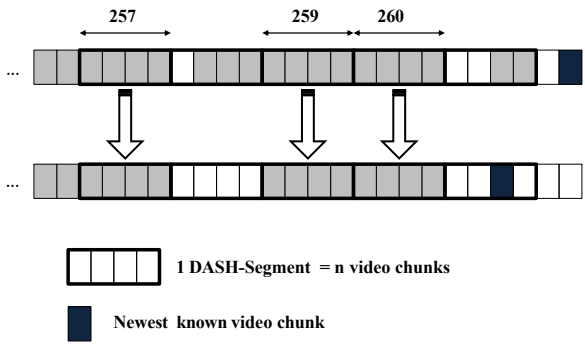}
\caption{DASH-segment inheritance  in migration process} 
\label{fig:cleaning-buffer}
\end{figure}

The last numerical investigation we perform is centered on DASH  and
on its availability to move from one representation to another with a time granularity which is equal to the size of a segment.
Namely, we want to assess how this very peculiar feature can be profitably employed in a P2P-based platform and for doing so we assume that 
when a peer moves from one overlay to another, it can
successfully inherit all DASH segments currently present within its buffer. This content will contribute to avoid video stalls 
and it will ease the transition to the new video representation.
Only the
completely received segments are inherited, and this statement is made clear by the example portrayed in Fig.\ref{fig:cleaning-buffer}, where the segment with sequence number $258$ is lost in the migration. 
To have a sound term of comparison, we consider the reference case where the node joining the new overlay  is forced to remove from its  buffer the video chunks of the representation it was previously receiving  and to collect the video chunks of the new representation from scratch. 
We then evaluate the switching delay, that we define as the time a  peer needs to gather $8$ s of consecutive video chunks every time it migrates from its current overlay to a new one and compute this delay from the time the peer joins the new overlay; we believe $8$ s to be a reasonable time that the peer has to wait before letting the playout begin. 
\begin{figure}[htbp]
\begin{minipage}[b]{1.0\linewidth}
\centering
\includegraphics[width= \columnwidth]{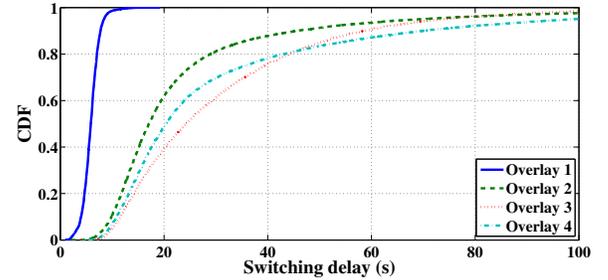}
 \centerline{(a) emptying the buffer }\medskip
\end{minipage}
\begin{minipage}[b]{1.0\linewidth}
\centering
\includegraphics[width= \columnwidth]{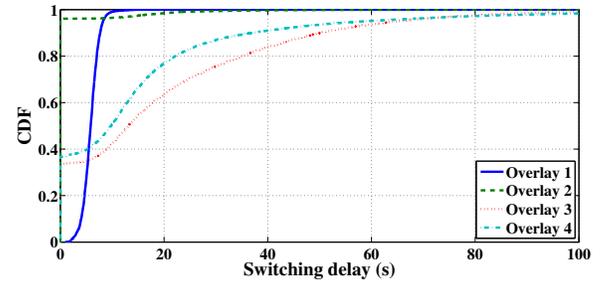}
 \centerline{(b) inheriting buffer content}\medskip
\end{minipage}
\caption{Switching delay CDF}
\label{fig:startup} 
\end{figure}
Fig.\ref{fig:startup}(a) reports the CDF of the switching delay in the different overlays, as numerically determined for the reference case.
This figure shows that the delay increases with increasing values of the video streaming rate, i.e., from overlay $1$ to $2$, $3$ and $4$
and quantitatively provides evidence that the choice to force every node to start from the lowest representation is correct,
as it statistically guarantees the lowest latency. 
%
%
Next, we consider the DASH-supported solution where the peer scans its buffer to evaluate if it can reuse some of its video content.
Fragments of DASH segments that might be present within the buffer at migration time are removed, as useless.
%
%
%
The obtained results are reported in Fig.\ref{fig:startup}(b) and refer to a duration of the DASH segment equal to $2$ s. 
%
From the comparison with Fig.\ref{fig:startup}(a)
we notice that the delay for overlay $1$ does not change, as it is modestly influenced by the migration of the peers; on the contrary, DASH guarantees a considerable reduction of the delay for the 
remaining overlays. 
As an impressive example,  the time 
needed to accumulate $8$ s of video when the peer moves to the overlay that distributes representation $2$ is $0$ s with probability $0.95$, meaning that with high probability a significant portion of the current buffer content is
profitably employed during the migration process towards overlay $2$. By visual inspection of the two figures we conclude that the same remark applies to the delay 
that the peer experiences to accumulate $8$ s when it moves to the overlays distributing representation $3$ and $4$. 

\section{Conclusions}
This paper has proposed a 
P2P-DASH architecture that jointly exploits the cooperation property of P2P and the flexibility of DASH to stream 
good quality videos
to a large population of users. Each peer within the system 
relies upon a decentralized rate control strategy
that steers its rate variations, hence its movements from one overlay to another, on the basis of local and global indicators reflecting the 
health status of the
single peer and of the system overlays.
The effectiveness of the proposed solution has been demonstrated through simulations, indicating that the P2P-DASH platform is able to guarantee its users a very good performance;
its overlays operate in much better conditions than the overlays of a conventional DASH-unaware architecture
subject to the same streaming requests and relying upon the same availability of peers' upload capacity. 
Moreover, through a comparison with a reference system modeled via an integer linear programming problem
it has been shown that our system also outperforms such reference architecture.
Finally, system behavior has been investigated in the critical condition of a flash crowd, demonstrating that the harsh and rapid input of a large number of new peers can be successfully revealed and gradually accommodated. 



%
%



%
\bibliographystyle{IEEEtran}
\bibliography{IEEEabrv,BIBLIO}

\end{document}